\documentclass[aps,prl,twocolumn,superscriptaddress,amsmath,showpacs]{revtex4-1}
\usepackage{epsfig}
\begin{document}

% Use the \preprint command to place your local institutional report
% number in the upper righthand corner of the title page in preprint mode.
% Multiple \preprint commands are allowed.
% Use the 'preprintnumbers' class option to override journal defaults
% to display numbers if necessary
%\preprint{}

%Title of paper
\title{Strongly Coupled Diamond Spin Qubits by Molecular Nitrogen Implantation}

% repeat the \author .. \affiliation  etc. as needed
% \email, \thanks, \homepage, \altaffiliation all apply to the current
% author. Explanatory text should go in the []'s, actual e-mail
% address or url should go in the {}'s for \email and \homepage.
% Please use the appropriate macro foreach each type of information

% \affiliation command applies to all authors since the last
% \affiliation command. The \affiliation command should follow the
% other information
% \affiliation can be followed by \email, \homepage, \thanks as well.
\author{Takashi Yamamoto}
\email[]{yamamoto.takashi@nims.go.jp, Current affiliation: National Institute for Materials Science, 1-1 Namiki, Tsukuba, Ibaraki 305-0044, Japan.}
\affiliation{Japan Atomic Energy Agency, 1233 Watanuki, Takasaki, Gunma 370-1292, Japan}
\author{Christoph M\"{u}ller}
\affiliation{Institute for Quantum Optics and Center for Integrated Quantum Science and Technology (IQ$^{\textit{st}}$), University of Ulm, D-89081, Ulm, Germany}
\author{Liam P. McGuinness}
\email[]{liam.mcguinness@uni-ulm.de}
\affiliation{Institute for Quantum Optics and Center for Integrated Quantum Science and Technology (IQ$^{\textit{st}}$), University of Ulm, D-89081, Ulm, Germany}
\author{Tokuyuki Teraji}
\affiliation{National Institute for Materials Science, 1-1 Namiki, Tsukuba, Ibaraki 305-0044, Japan}
\author{Boris Naydenov}
\affiliation{Institute for Quantum Optics and Center for Integrated Quantum Science and Technology (IQ$^{\textit{st}}$), University of Ulm, D-89081, Ulm, Germany}
\author{Shinobu Onoda}
\affiliation{Japan Atomic Energy Agency, 1233 Watanuki, Takasaki, Gunma 370-1292, Japan}
\author{Takeshi Ohshima}
\affiliation{Japan Atomic Energy Agency, 1233 Watanuki, Takasaki, Gunma 370-1292, Japan}
\author{J\"org Wrachtrup}
\affiliation{3rd Physics Institute and Research Center SCoPE, University of Stuttgart, D-70174, Stuttgart, Germany}
\author{Fedor Jelezko}
\affiliation{Institute for Quantum Optics and Center for Integrated Quantum Science and Technology (IQ$^{\textit{st}}$), University of Ulm, D-89081, Ulm, Germany}
\author{Junichi Isoya}
\affiliation{Research Center for Knowledge Communities, University of Tsukuba, 1-2 Kasuga, Tsukuba, Ibaraki 305-8550, Japan}

%\email[]{liam.mcguinness@uni-ulm.de}
%\homepage[]{Your web page}
%\thanks{}
%\altaffiliation{}

%Collaboration name if desired (requires use of superscriptaddress
%option in \documentclass). \noaffiliation is required (may also be
%used with the \author command).
%\collaboration can be followed by \email, \homepage, \thanks as well.
%\collaboration{}
%\noaffiliation

\date{\today}

\begin{abstract}
Ionized nitrogen molecules ($^{15}$N$_{2}^+$) are used as efficient point sources for creating NV$^-$ pairs in diamond with nanoscale spatial separation and up to 55~kHz magnetic coupling strength. Co-implantation of $^{12}$C$^+$ increased the yield of pairs, and a $^{13}$C-depleted diamond allowed 0.65~ms coherence times to be obtained. Further coupling to a third dark spin provided a strongly coupled three spin register. These results mark an important step towards realization of multi-qubit systems and scalable NV$^-$ quantum registers.
\end{abstract}

% insert suggested PACS numbers in braces on next line
%{Color centers and other defects, Ion radiation effect, Decoherence; open systems; quantum statistical methods}
% insert suggested keywords - APS authors don't need to do this
%\keywords{quantum information, qubit, magnetic coupling, diamond, NV center}

%\maketitle must follow title, authors, abstract, \pacs, and \keywords
\maketitle
The realization of cost effective poly-qubit registers has the potential to change modern society. Quantum processors are expected to outperform the best known algorithms operating on classical machines, allowing certain untenable problems to be carried out, like factorization of large numbers. Some solutions would undermine the current basis of information security, while others would extend the reach of computational methods to applications such as quantum chemistry and quantum simulation allowing a greater level of understanding of nature. However, quantum processors are yet to achieve this vast potential, and this is especially true for the most attractive of computing architectures -- solid state and near ambient conditions.

Spins of negatively-charged nitrogen-vacancy (NV$^-$) defects in diamond are amongst the leading solid-state quantum bits (qubits) operating under ambient conditions. The NV$^-$ defect consists of a substitutional nitrogen atom and an adjacent vacancy in the diamond lattice. An additional captured electron gives the defect a negative overall charge, resulting in a bright, single photon emitter with zero phonon line at 637~nm~\cite{Davies}. The $|{\it{m}}_{\rm{s}}=0\rangle$ and $|{\it{m}}_{\rm{s}}=\pm1\rangle$ sublevels of the ground-state spin triplet ($S=1$) are separated by $\sim$2.87~GHz due to spin-spin interaction~\cite{Reddy}. The ground states can be optically initialized, manipulated, and then readout~\cite{Jelezko1,Jelezko2}. In complement to these characteristics is an extraordinarily long spin coherence time, over a millisecond at room temperature (RT)~\cite{Balasubramanian, Ishikawa, Jahnke}.

For quantum technology applications such as quantum information processing, the challenge remains to create series of coupled NV$^-$ centers which may form the processor of the quantum computer. One avenue to creating scalable quantum registers involves fabrication of an array of NV$^-$ centers with nanometer separations. In this scheme, adjacent NV$^-$ centers are coupled by magnetic dipole interaction, which scales as the inverse cube of separation distance~\cite{Neumann}. In addition to strong dipolar coupling among single NV$^-$ spins, long spin coherence times are essential, with the preferred route to building such a qubit network being position controlled ion implantation into pure diamond~\cite{Neumann,Meijer,Gaebel,Rabeau,Naydenov1,Naydenov2,Pezzagna1,Schwartz,Pezzagna2,Pezzagna3,Toyli,Dolde}. However obstacles such as reduced coherence times due to implantation damage and insufficient spatial accuracy have hindered results. To date, dipolar coupling~\cite{Neumann} and entanglement~\cite{Dolde} between two NV$^-$ qubits has been shown, but the short coherence times, below or on the order of the coupling rate limited the fidelity of two qubit gates. A further limitation has been the extremely low yield ($\sim0.1\%$) of generating NV$^-$ pairs~\cite{Gaebel,Toyli}.

Imperfect conversion of implanted nitrogen to NV$^-$ is a crucial obstacle to achieving multi-qubit systems since the probability to create a pair of NVs goes as the square of the creation efficiency, while for three and four qubit arrays the scaling is the yield cubed and to the power of four respectively, meaning achievement of high yields is especially important. Nevertheless, the reliable generation of two coupled NV$^-$ centers would be significant progress, since two strongly coupled NV$^-$ qubits, together with their intrinsic nuclear spins would, for example allow quantum error correction protocols, high resolution gradient magnetometry~\cite{Shin}, or entanglement assisted magnetometry to be performed in the solid state. Here we obtain a 4\% yield of pairs, a substantial improvement on previous results. We also report the first observation of a strongly coupled $^{15}$NV$^-$ pair, with a coupling strength, $J$, exceeding the inhomogenous linewidth (1/{\it{T}}$_{2}^{\>*}$) by more than a factor of five. The product of the coherence time ($T_2$) and the coupling rate can be used to give a measure of gate fidelity. By comparing the factor that we obtain ($T_2\times J = 36$), to other architectures, we find it is comparable to ion traps ($T_2\times J \approx 34$)\cite{Gulde,Leibfried} and superior to superconducting qubits ($T_2\times J \approx 10$)\cite{DiCarlo}, placing this coupled quantum system amongst the leading in any architecture.

\begin{figure}
\includegraphics[width=8.5cm]{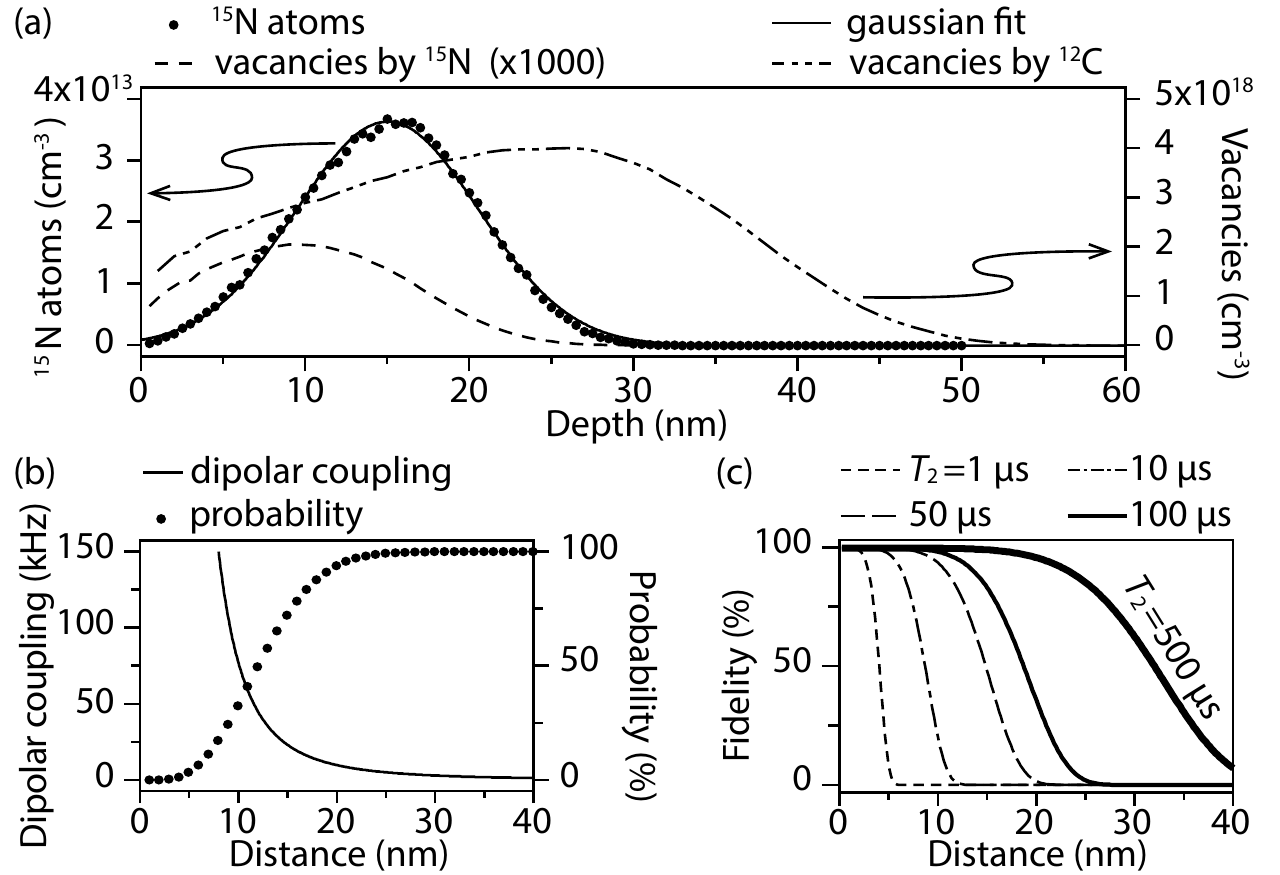}%
\caption{{\textbf{Simulation of spatial distribution for $^{15}$N ions and vacancies.}} (a) Simulated depth distribution of implanted $^{15}$N atoms with 10~keV energy (solid circles), Gaussian fit (solid line), distribution of vacancies arising from $^{15}$N implantation (dashed line), and distribution of vacancies from 20~keV $^{12}$C co-implantation (dot-dashed line). (b) Simulated probability distribution of distance between two $^{15}$N atoms arising from the same N$_2$ implanted molecule (solid circles), and the resultant dipolar coupling between two NV$^-$ centers with this separation (solid line). Note that we averaged the angular component in the dipolar coupling of $\nu_\text{dip}=\frac{3}{2}D_\text{dip}\overline{|3\cos^2\theta-1|}$  where $\theta$ is the angle between the vector {\textbf{r}} connecting two dipoles and the magnetic field, and the dipolar coupling constant $D_\text{dip}\sim5.2$$\times10^{-23}/r^{3}$~kHz. (c) Calculated fidelity of $\exp[-(\frac{1}{\nu_\text{dip}T_2})^2]$ for a two-qubit entanglement with various coherence times $T_2$.\label{FIG1}}
\end{figure}

The sample used in this study was a high-purity (nitrogen concentration $<1$~ppb) and isotopically-purified ($^{12}$C-99.998\%) homoepitaxial (100) diamond film grown by microwave plasma-assisted chemical vapor deposition~\cite{Teraji}. Ionised nitrogen molecules, $^{15}$N$_2^+$, were implanted into the diamond with an acceleration voltage of 20~keV. Afterwards $^{12}$C$^+$ ions with 20~keV energy were co-implanted into one half of the $^{15}$N$_2$ implant region. The ion fluences were 2.5$\times10^7$~$^{15}$N$_2^+$/cm$^2$ and 1.4$\times10^{11}$~$^{12}$C$^+$/cm$^2$, respectively. The simulated distribution of implanted ions and vacancies as computed by SRIM code (Stopping and Range of Ions in Matter/ver. 2008~\cite{Ziegler}) is shown in Figure~\ref{FIG1}(a). The average depth of individual $^{15}$N ions with 10~keV acceleration voltage (20~keV per molecule) is calculated as 15~nm. The vacancy distribution due to $^{12}$C implantation peaks at $\sim$25~nm, but contributes approximately 2,000 times more vacancies than the $^{15}$N$_2$ implantation (dashed line) over the stopping range of the nitrogen ions. This allows NV$^{-}$ formation with a high yield~\cite{Naydenov1} and the conversion of $^{15}$N$_2$ ions to $^{15}$NV-$^{15}$NV pairs with an increased probability when compared to previous studies at similar energies. Previous methods to achieve coupling between two NV$^-$ centers involved using focused beams~\cite{Neumann} or pinhole apertures~\cite{Dolde}. Here, molecular $^{15}$N$_2^+$ implantation was used as an ultimate point source~\cite{Gaebel}, where the distance between two N atoms from the same molecule is determined only by the ion straggling length. The in-depth $^{15}$N ion straggling as shown in Figure~\ref{FIG1}(a) has a width (2$\sigma$) of 11.1~nm, and the in-plane straggle has a calculated width of 8.9~nm (data not shown), where $\sigma$ is the standard deviation of the Gaussian fit. However we note that channeling of implanted nitrogen has not been taken into account may lead to increased separations. From this, the probability distribution for the spatial separation of two N atoms (Figure~\ref{FIG1}(b) solid circles) and the corresponding dipolar coupling for an NV$^-$ pair (black line) can be estimated. For a perfect NV$^-$ creation efficiency this gives a 41\% probability to produce a pair with separation less than 11~nm and greater than 59~kHz coupling strength from each implanted $^{15}$N$_2^+$ molecule. Figure~\ref{FIG1}(c) shows the calculated fidelity for a two-qubit entanglement with various coherence times, $T_2$. For coherence times of $100~\mu$s or more, a fidelity above 97\% can be obtained with this coupling strength (59~kHz).

To form NV configurations, the implanted sample was annealed at 1000$^{\circ}$C for 2~hours in vacuum. The resultant NV centers were measured using a home-built confocal microscope~\cite{Gruber} and the spin properties were observed through optically detected magnetic resonance (ODMR) spectroscopy~\cite{Rabeau}. Observation of either the implanted $^{15}$N hyperfine structure (with nuclear spin, $I=1/2$) or native $^{14}$N (I=1, natural abundance 99.63\%) by ODMR spectroscopy allowed determination of whether the investigated NV$^-$ centers were due to implantation or pre-existing impurities in the substrate~\cite{Rabeau}.
\begin{figure}
\includegraphics[width=8.5cm]{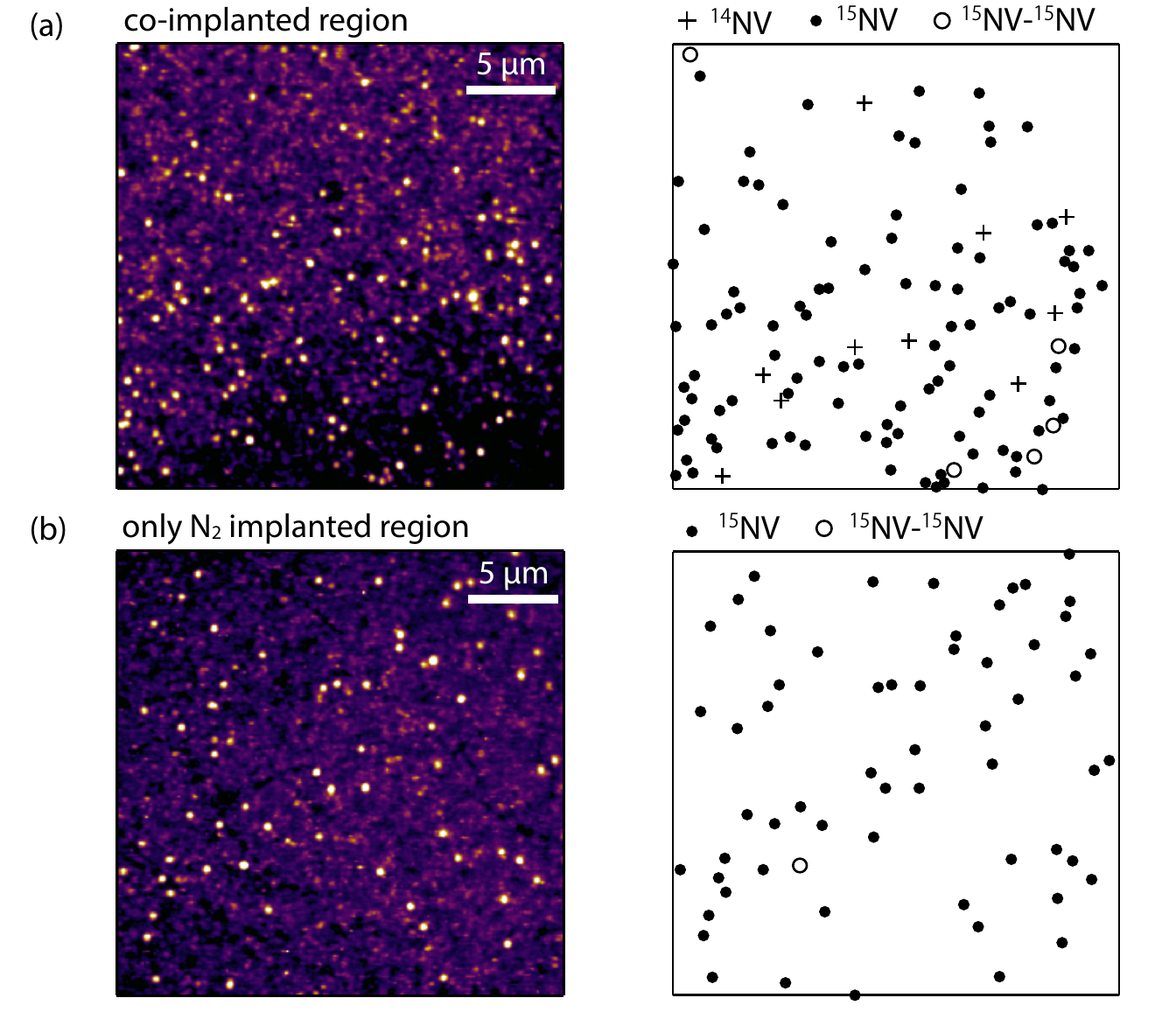}%
\caption{(Color online) {\textbf{Confocal microscopy of NV$^-$ centers.}} (a) A fluorescence confocal image of the co-implanted ($^{15}$N$_2$ and $^{12}$C implanted) region with an area of 25$\times$25~$\mu$m$^2$ ({\it{left}}). Bright yellow dots correspond to NV centers. The coordinate mapping of ODMR identified $^{14}$NV$^-$ (cross), $^{15}$NV$^-$ (solid circle), and $^{15}$NV$^-$-$^{15}$NV$^-$ pairs (open circle) is shown in the right figure. (b) A fluorescence confocal image of the $^{15}$N$_2$-only implanted region with an area of 25$\times$25~$\mu$m$^2$ (left) and the coordinate mapping (right) showing $^{15}$NV$^-$ (solid circle) and an $^{15}$NV$^-$-$^{15}$NV$^-$ pair (open circle).
\label{FIG2}}
\end{figure}

A confocal image of the co-implanted region is shown in Figure~\ref{FIG2}(a). ODMR spectroscopy revealed that 100 $^{15}$NV$^-$ centers, five $^{15}$NV$^-$-$^{15}$NV$^-$ pairs, and 10 $^{14}$NV$^-$ centers were created in a 25$\times$25~$\mu$m$^2$ co-implanted area, from $\sim$156 implanted N$_2$ molecules. As ODMR spectroscopy is unable to resolve NV$^-$ pairs with the same orientation, and this is expected to occur in one quarter of pairs (due to the NV$^-$ $C_{3v}$ symmetry), the number of pairs will be underestimated by a factor of 0.25. Therefore we estimate the $^{15}$NV$^-$ pair creation efficiency per implanted molecule as 4$\pm$2\% ($6/156\approx0.04$). Counting the total number of observed $^{15}$NV$^-$ centers and dividing by the number of implanted ions in the 25$\times$25~$\mu$m$^2$ area gives a creation efficiency of 36$\pm$9\%. Comparison to a previous study of 60~keV $^{15}$N$_2^+$ and 40~keV $^{12}$C implantation with a 33\% creation yield, gives good agreement to the 36$\pm$9\% obtained in this study~\cite{Naydenov1}. In order to demonstrate the efficacy of co-implantation for creating NV$^-$ centers, a confocal map of the N$_2$-only implanted region was also performed (see Figure~\ref{FIG2}(b)). In this  25$\times$25~$\mu$m$^2$ area, 62 single $^{15}$NV$^-$ centers and a lone $^{15}$NV$^-$-$^{15}$NV$^-$ pair were produced, giving a creation efficiency of 20$\pm$7\%. We find that co-implantation approximately doubles the NV$^-$ creation efficiency and improves NV$^-$ pair production by a factor of four when compared to N$_2$-only implantation (see also Ref.~\onlinecite{Naydenov1}).

\begin{figure}
\includegraphics[width=8.5cm]{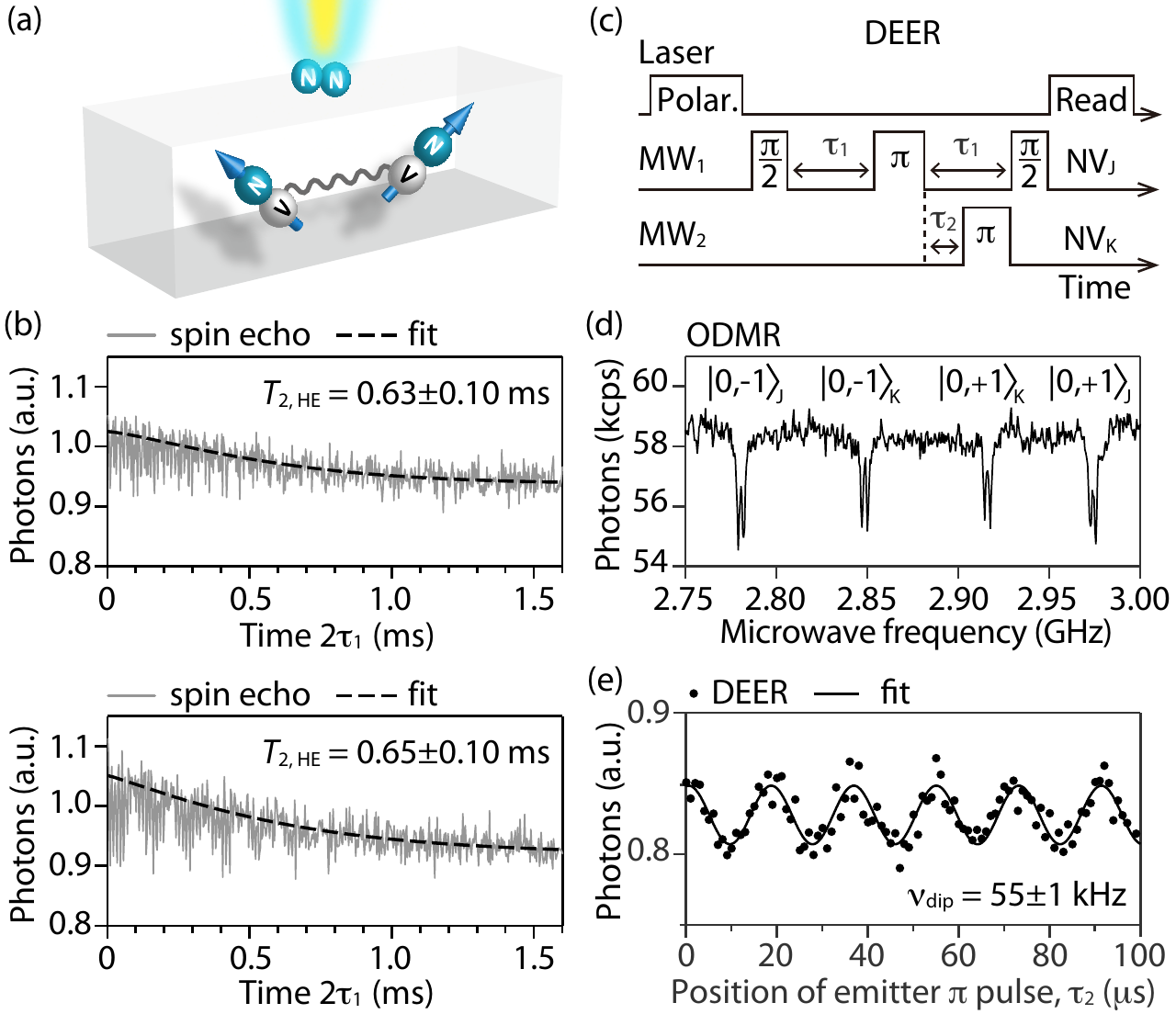}%
\caption{(Color online) {\textbf{Dipolar coupling of a $^{15}$NV$^-$ pair measured by DEER.}} (a) Illustration of NV pair creation by N$_2^+$ molecular ion implantation. (b) Measured spin echo data for the $^{15}$NV$^-$ pair are fit with $E(2\tau)=a+b\exp[-(2\tau/T_{2})^\alpha]$ showing long $T_{2, {\text{HE}}}$ times of 0.65~ms ($\alpha=1.14$) and 0.63~ms ($\alpha=1.31$). (c) DEER pulse sequence with NV$_{\textrm{J}}$ as sensor and NV$_{\textrm{K}}$ as emitter. (d) ODMR spectrum of the investigated NV pair shows the selective addressability of the two $^{15}$NV$^-$ centers due to their different axis orientation, where a magnetic field of $\sim34$~G parallel to NV$_{\textrm{J}}$ axis. (e) The resulting echo modulation on NV$_{\textrm{J}}$ shows a coupling strength of 55~kHz.\label{FIG3}}
\end{figure}

Now we turn our attention to a particular coupled $^{15}$NV$^-$ pair shown schematically in Figure~\ref{FIG3}(a). Hahn echo was performed on each NV$^-$~\cite{Gaebel} and the decay curves were fitted by $E(2\tau)\propto\exp[-(2\tau/T_{2})^\alpha]$ where $\alpha$ is a free parameter (Figure~\ref{FIG3}(b)). The long coherence times of 0.63$\pm$0.10 and 0.65$\pm$0.10~ms are, to the best of our knowledge, a record for shallow implanted NV$^-$ centers, a significant improvement on the 0.35~ms recorded for 14~keV-$^{14}$N$_2$ implantation into 1.1\% $^{13}$C diamond~\cite{Gaebel}, and 0.15 and 0.5~ms for a coupled pair by 1~MeV-$^{15}$N implantation into 0.01\% $^{13}$C diamond~\cite{Dolde}. The exponent of the Hahn echo decay is related to the fluctuation regime of magnetic noise in the environment~\cite{Hall}. An exponent close to unity as observed here ($\alpha=1.14$ and 1.31), indicates that fast noise from electron spins on surface or residuals surrounding the NV$^-$ is a major source of decoherence. Thus, there may be room to improve the coherence times by overgrowth on the ion implanted surface in order to remove surface noise~\cite{Staudacher}. We also note that a Gaussian or quartic exponent is beneficial for performing high fidelity operations as opposed to exponential decay observed for ion trap qubits encoded in the metastable state.

The truncated spin Hamiltonian of an NV$^-$ center ($S_{\textrm{J}}$) coupled to another NV$^-$ center ($S_{\textrm{K}}$) with gyromagnetic ratio $\gamma_e$ can be written as $\hat{H}/h=D\hat{S}_{\textrm{J}}^{2}+\gamma{_e}\hat{S}_{\textrm{J}}\Vec{B}_z+\hat{S}_{\textrm{J}}\hat{A}\hat{I}_{\textrm{J}}+\nu_{\textrm{dip}}\hat{S}_{\textrm{J}}\hat{S}_{\textrm{K}}$ where the first term describes the zero field splitting ($D=2.87$~GHz), the second term describes Zeeman splitting in an applied field $\Vec{B}_z$, the third term is the hyperfine coupling ($A=3.1$~MHz for $^{15}$N), and the last term is the magnetic dipolar coupling between spins with coupling frequency $\nu_{\textrm{dip}}$. To investigate the magnetic coupling between the pair, we employed a double electron-electron resonance (DEER) technique (Figure~\ref{FIG3}(c)). We note that this technique is possible due to the different orientation of the two NV centers in an applied magnetic field, allowing them to be independently addressed by microwaves (Figure~\ref{FIG3}(d)). As can be seen in Figure~\ref{FIG3}(e), periodic modulation with a dipolar coupling frequency of $\nu_{\textrm{dip}}=55$~kHz between the NV pair occurs. The low decoherence over the timescale of this coupling as observed by the prolonged contrast of the modulations should allow high fidelity entanglement between the qubits. From the magnetic coupling strength and the coherence time, we obtain a maximum expected fidelity for 2-qubit entanglement in excess of 99.9\%, in good comparison to the best experimental results with ion qubits \cite{Gulde}.
\begin{figure}
\includegraphics[width=8.6cm]{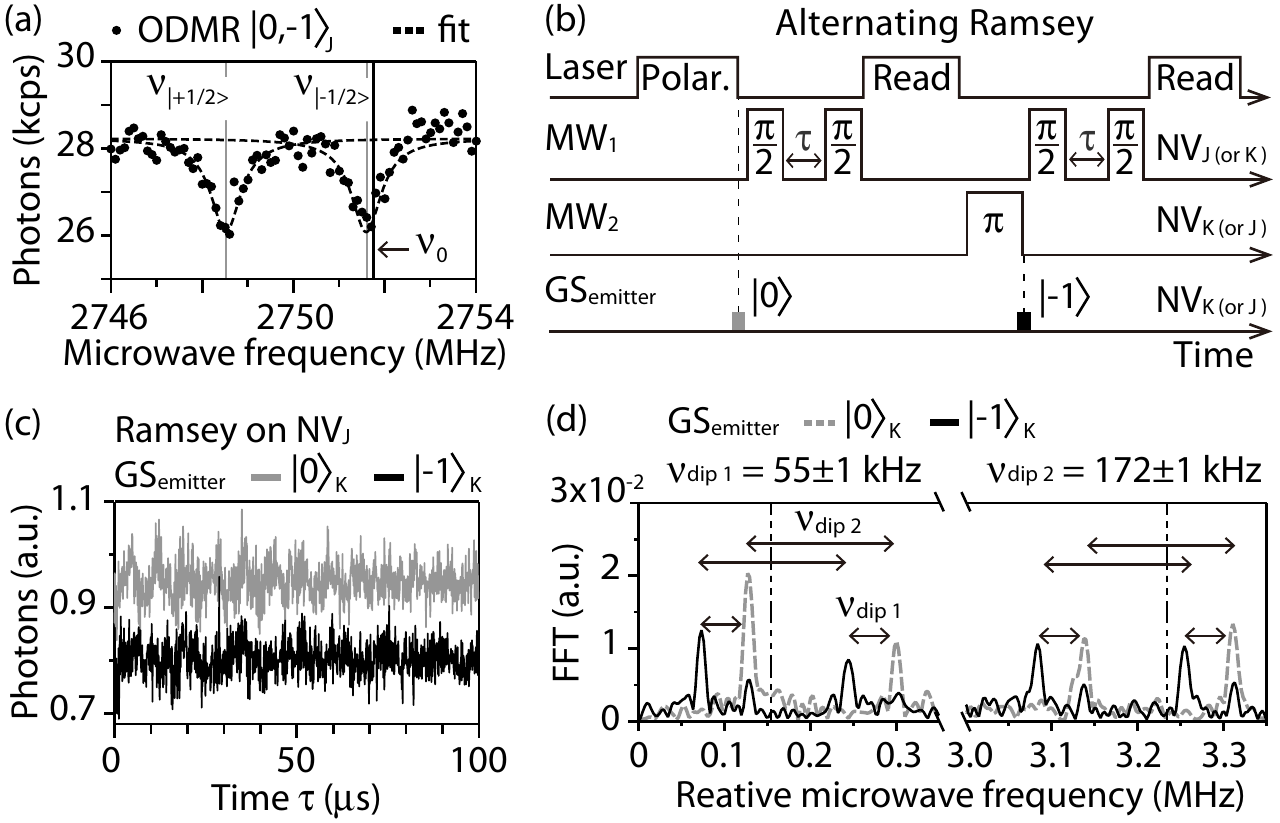}%
\caption{{\textbf{Ramsey measurements on a coupled $^{15}$NV$^-$ pair and third dark spin.}} (a) Hyperfine spin transitions between $m_{\rm{s}}=0, -1$ states for NV$_{\textrm{J}}$. A magnetic field of $\sim46$~G was applied parallel to NV$_{\textrm{J}}$. The detuning frequency ($\nu_0$) used for the $\pi/2$ pulses in the Ramsey sequence is indicated by the arrow. (b) The alternating Ramsey sequence. Two different ground states, GS$_\textrm{emitter}$, of $|0\rangle$ and $|-1\rangle$ are prepared without and with a $\pi$ pulse on the emitter spin, respectively. (c) Experimental data of the Ramsey pulse sequence, showing a $T_2^{\>*}$ for NV$_{\textrm{J}}$ of 100~$\mu$s. A vertical offset for the data with (black) and without (gray) the $\pi$ pulse on the emitter spin was added for pictorial clarity. (d) The FFT of the obtained Ramsey data, showing a shift of 55~kHz for the two different states of the emitter NV and also an additional splitting from the expected frequencies (dot-dashed lines) of 172~kHz.\label{FIG4}}
\end{figure}

Coupling between the $^{15}$NV$^-$ pair was also investigated by Ramsey spectroscopy. The microwave frequency $\nu_0$ was slightly detuned from a hyperfine transition of the sensor spin, NV$_{\textrm{J}}$ (see Figure~\ref{FIG4}(a)), and two $\pi/2$ pulses separated by a delay of $\tau$ were applied. This sequence was interleaved with another identical sequence differing only by the application of a $\pi$ pulse on the emitter spin, NV$_{\textrm{K}}$ at the beginning of the sequence (Figure~\ref{FIG4}(b)). Ramsey spectroscopy also yields an oscillating output (Figure~\ref{FIG4}(c)), where now the oscillation frequencies are simply the detuning of the $\pi/2$ pulses from the natural frequencies of the system. The $\pi$ pulse on the emitter spin shifts these frequencies by the coupling strength, which can be directly observed by comparing the Fast Fourier transform (FFT) power spectrum of the measured Ramsey fringes, with and without the selective $\pi$ pulse, giving a value of 55$\pm$1~kHz for the coupling strength (Figure~\ref{FIG4}(d)). Interestingly, we also observe another pair of frequencies in the power spectrum, where each hyperfine transition is split by 172$\pm$1~kHz. One may attribute this to electric interaction on the sensor spin, occurring when photoionization of the emitter spin switches its charge state between NV$^-$ and NV$^0$. However calculations based on the electric field sensitivity of NV$^-$ to a single electric charge gives a maximal expected separation of approximately 2~nm, much less than that obtained through the magnetic coupling. In addition, the $\pi$ pulse on the emitter spin, is conditional on the NV being in a negative charge state, therefore we would not expect a 55~kHz shift when the emitter is in the NV$^0$ charge state. Instead we attribute the additional splitting as being due to magnetic coupling with a third dark spin, situated between NV$_{\textrm{J}}$ and NV$_{\textrm{K}}$, providing a strongly coupled three spin register. By changing the Ramsey sequence to now use NV$_{\textrm{K}}$ to sense the magnetic field emitted by NV$_{\textrm{J}}$, we again observe a frequency shift of 53$\pm$3~kHz due to the magnetic dipolar coupling and an additional splitting of 330$\pm$2~kHz (data not shown), showing that the dark spin is coupled more strongly to NV$_{\textrm{K}}$.

Finally, by integrating over remnant peaks in the FFT spectrum, the Ramsey experiment also allows us to probe the initialization and pulse fidelity of the NV pair. The gray dashed curve in Figure~\ref{FIG4}(d), has no secondary peaks 55~kHz from the main peaks, indicating almost perfect spin initialization, whereas the black solid curve (after application of a $\pi$ pulse on NV$_{\textrm{K}}$) displays secondary peaks at 55~kHz with 30-35\% the intensity of the main peaks, indicating the spin manipulation is imperfect, with a fidelity of 65-70\%. Performing the same analysis on the data with NV$_{\textrm{K}}$ as sensor and NV$_{\textrm{J}}$ as emitter, yielded almost perfect spin initialization and manipulation with more than 90\% fidelity.

In summary, engineered NV qubits were introduced into a spin free lattice of $^{13}$C-depleted diamond by co-implantation of low energy nitrogen molecules and $^{12}$C$^+$ ions, and high temperature annealing. An improved $^{15}$NV$^-$ creation efficiency of $\sim$36\% for single centers, and 4\% for pairs was demonstrated, of which half are expected to have a coupling strength exceeding 45~kHz. We reported on the creation of one such $^{15}$NV$^-$ pair with a magnetic dipolar coupling of 55~kHz, and spin coherence time of more than 0.6~ms. A further coupling of 172 and 330~kHz between the pair and a third dark spin was observed. Using low energy molecular $^{15}$N$_2^+$ implantation as a nearly ideal point source, we have thus demonstrated an efficacious route to achieving two NV qubit systems with long coherence times. The improved creation of NV$^-$ centers shown here is an important step towards the scalability of quantum registers created by implanting N$_3$ and N$_4$ molecules~\cite{Bieske} and can also be applied to phosphorus in silicon architectures~\cite{Wilson}.

This study was carried out as {\textquoteleft}Strategic Japanese-German Joint Research Project' supported by JST and DFG (FOR1482, FOR1493, SFB/TR 21), EU (DIAMANT), and the Alexander von Humboldt Foundation.


\begin{thebibliography}{99}
\bibitem{Davies} G. Davies and M. F. Hamer, Proc. R. Soc. London Ser. A {\textbf348}, 285 (1976).
\bibitem{Reddy} N. R. S. Reddy, N. B. Manson, E. R. Krausz, J. Lumi. {\textbf38}, 46 (1987).
\bibitem{Jelezko1} F. Jelezko, T. Gaebel, I. Popa, M. Domhan, A. Gruber, and J.Wrachtrup, Phys. Rev. Lett. \textbf{93}, 130501 (2004).
\bibitem{Jelezko2} F. Jelezko and J. Wrachtrup, Phys. Stat. Sol. {\textbf(a) 203}, 3207 (2006).
\bibitem{Balasubramanian} G. Balasubramanian, P. Neumann, D. Twitchen, M. Markham, R. Kolesov, N. Mizuochi, J. Isoya, J. Achard, J. Beck, J. Tisler, V. Jacques, P. R. Hemmer, F. Jelezko, and J. Wrachtrup, Nat. Mat. \textbf{8}, 383 (2009).
\bibitem{Ishikawa}T. Ishikawa, K. C. Fu, C. Santori, V. M. Acosta, R. G. Beausoleil, H. Watanabe, S. Shikata, and K. M. Itoh, Nano Lett. {\textbf{12}}, 2083 (2012).
\bibitem{Jahnke} K.D. Jahnke, B. Naydenov, T. Teraji, S. Koizumi, T. Umeda, J. Isoya, and F. Jelezko, Appl. Phys. Lett. {\textbf{101}}, 012405 (2012).
\bibitem{Neumann} P. Neumann, R. Kolesov, B. Naydenov, J. Beck, F. Rempp, M. Steiner, V. Jacques, G.
Balasubramanian, M. L. Markham, D. J. Twitchen , S. Pezzagna, J. Meijer, J. Twamley, F.Jelezko and J. Wrachtrup, Nat. Phys. \textbf{6}, 249 (2010).
\bibitem{Meijer} J. Meijer, B. Burchard, M. Domhan, C. Wittmann, T. Gaebel, I. Popa, F. Jelezko, and J. Wrachtrup, Appl. Phys. Lett. \textbf{87}, 261909 (2005).
\bibitem{Gaebel} T. Gaebel, M. Domhan, I. Popa, C. Wittmann, P. Neumann, F. Jelezko, J. R. Rabeau, N. Stavrias, A. D. Greentree, S. Prawer, J. Meijer, J. Twamley, P. R. Hemmer, and J. Wrachtrup, Nat. Phys. \textbf{2}, 408 (2006).
\bibitem{Rabeau} J. R. Rabeau, P. Reichart, G. Tamanyan, D. N. Jamieson, S. Prawer, F. Jelezko, T. Gaebel, I. Popa, M.Domhan, and J. Wrachtrup, Appl. Phys. Lett. \textbf{88}, 023113 (2006).
\bibitem{Naydenov1} B. Naydenov, V. Richter, J. Beck, M. Steiner, P. Neumann, G. Balasubramanian, J. Achard, F. Jelezko, J. Wrachtrup, and R. Kalish, Appl. Phys. Lett. \textbf{96}, 163108 (2010).
\bibitem{Naydenov2} B. Naydenov, F. Reinhard, A. L\"{a}mmle, V. Richter, R. Kalish, U. F. S. D'Haenens-Johansson, M. Newton, F. Jelezko, and J. Wrachtrup, Appl. Phys. Lett. \textbf{97}, 242511 (2010).
\bibitem{Pezzagna1} S. Pezzagna, B. Naydenov, F. Jelezko, J. Wrachtrup, and J. Meijer, New J. Phys. \textbf{12}, 065017 (2010).
\bibitem{Pezzagna2} S. Pezzagna, D. Wildanger, P. Mazarov, A. D. Wieck, Y. Sarov, I. Rangelow, B. Naydenov, F. Jelezko, S. W. Hell, and J. Meijer, Small \textbf{6}, 2117 (2010).
\bibitem{Toyli} D. M. Toyli, C. D. Weis, G. D. Fuchs, T. Schenkel, and D. D. Awschalom, Nano Lett. \textbf{10}, 3168 (2010).
\bibitem{Schwartz} J. Schwartz, P. Michaelides, C. D. Weis, T. Schenkel, New Journal of Physics \textbf{13}, 035022 (2011).
\bibitem{Pezzagna3} S. Pezzagna, D. Rogalla, H.-W. Becker, I. Jakobi, F. Dolde, B. Naydenov, J. Wrachtrup, F. Jelezko, C. Trautmann, and J. Meijer, Phys. Status Solidi A \textbf{208}, 2017 (2011).
\bibitem{Dolde} F. Dolde, I. Jakobi, B. Naydenov, N. Zhao, S. Pezzagna, C. Trautmann, J. Meijer, P. Neumann, F. Jelezko, and J. Wrachtrup, Nat. Phy.  \textbf{9}, 139 (2013)
\bibitem{Shin} C. Shin, C. Kim, R. Kolesov, G. Balasubramanian, F. Jelezko, J. Wrachtrup, and P. R. Hemmer, J. Lumi. \textbf{130}, 1635 (2010).
\bibitem{Gulde} S. Gulde, M. Riebe, G. P. T. Lancaster, C. Becher, J. Eschner, H. H\"affner, F. Schmidt-Kaler, I. L. Chuang, and R Blatt, Nature \textbf{421}, 48 (2003).
\bibitem{Leibfried} D. Leibfried, B. DeMarco, V. Meyer, D. Lucas, M. Barrett, J. Britton, W. M. Itano, B. Jelenkovi\'c, C. Langer, T. Rosenband, and D. J. Wineland, Nature \textbf{422}, 412 (2003).
\bibitem{DiCarlo} L. DiCarlo, J. M. Chow, J. M. Gambetta, Lev S. Bishop, B. R. Johnson, D. I. Schuster, J. Majer, A. Blais, L. Frunzio, S. M. Girvin, and R. J. Schoelkopf, Nature \textbf{460}, 240 (2009).
\bibitem{Teraji} T. Teraji, T. Taniguchi, S. Koizumi, K. Watanabe, M. Liao, Y. Koide, and J. Isoya, Jpn. J. Appl. Phys. \textbf{51}, 090104 (2012).
\bibitem{Ziegler} J. F. Ziegler, \emph{The Stopping and Ranges of Ions in Matter}, http://www.srim.org/
\bibitem{Gruber} A. Gruber, A. Dr{\"{a}}benstedt, C. Tietz, L. Fleury, J. Wrachtrup, and C. von Borczyskowski, Science \textbf{276}, 2012 (1997).
%\bibitem{Koike} J. Koike, D. M. Parkin and T. E. Mitchell, Appl. Phys. Lett. \textbf{60}, 1450 (1992).
%\bibitem{INSPEC} R. Kalish, 1994, \emph{Properties and Growth of Diamond}, edited by G. Davies, (INSPEC, London), p.~193, Refs therein.
%\bibitem{Saada} D. Saada, J. Adler, and R. Kalish, Int. J. Mod. Phys. C \textbf{09}, 61 (1998).
\bibitem{Hall} L. T. Hall, J. H. Cole, C. D. Hill, and L. C. L. Hollenberg, Phys. Rev. Lett. \textbf{103}, 220802 (2009).
\bibitem{Staudacher} T. Staudacher, F. Ziem, L. H\"{a}ussler, R. St\"{o}hr, S. Steinert, F. Reinhard, J. Scharpf, A. Denisenko, and J. Wrachtrup, Appl. Phys. Lett. \textbf{101}, 212401 (2012).
\bibitem{Bieske} E. Bieske, J. Chem. Phys. \textbf{98}, 8537 (1993).
\bibitem{Wilson} H. F. Wilson, S. Prawer, P. G. Spizzirri, D. N. Jamieson, N. Stavrias, D. R. McKenzie, Nucl. Instrum. Methods Phys. Res. B \textbf{251}, 395 (2006).

\end{thebibliography}
\end{document}